\documentstyle[epsfig,epsf]{article}
\newcommand{\ndt}{\noindent}

\def\nmne {\nu_\mu \longleftrightarrow \nu_e}
\def\nmnt {\nu_\mu \longleftrightarrow \nu_\tau}
\def\nmns {\nu_\mu \longleftrightarrow \nu_{sterile}}

\def\nm {\nu_\mu}

\def\Dm2{\Delta m^2}
\def\s2t{\sin^2 2\theta}
\def\Dv{\Delta v}
\def\aprle{\buildrel < \over {_{\sim}}}

\def\ll{\langle}
\def\rr{\rangle}

\def\be{\begin {equation}}
\def\ee{\end {equation}}

\begin{document}

\begin{center}
{\Large {\bf Search for a Lorentz invariance violation in 
atmospheric neutrino oscillations using MACRO data}}
\end{center}

\vskip .7 cm

\begin{center}
Miriam Giorgini \par~\par
{\it Dept of Physics, Univ. of Bologna and INFN, \\
V.le C. Berti Pichat 6/2, Bologna, I-40127, Italy\\} 

E-mail: miriam.giorgini@bo.infn.it

\par~\par
Invited talk at C2CR 2005, From Colliders to Cosmic Rays, \\
Prague, Czech Republic, 7-13 September 2005. 

\vskip .7 cm
{\large \bf Abstract}\par
\end{center}

{\small
The energy spectrum of neutrino-induced upward-going muons in MACRO was 
analysed in terms of special relativity principles violating 
effects, keeping standard 
mass-induced atmospheric neutrino oscillations as the dominant source of 
$\nmnt$ transitions. The data disfavour these exotic
possibilities even at a sub-dominant level, and stringent $90\%$ C.L. 
limits are placed on the Lorentz invariance violation parameters.
These limits can also be re-interpreted as upper bounds on the
parameters describing violation of the Equivalence Principle.}

\vspace{5mm}

\section{Introduction}\label{sec:intro}
The phenomenon of neutrino flavour oscillations, induced by flavour-mass
eigenstate mixing, is now considered a solid explanation of both solar 
and atmospheric neutrino data 
\cite{solar,macro-98,macro-last,sk-general,soudan2}.
In particular, the deficit of atmospheric muon neutrinos is well described 
by the two flavour $\nmnt$ oscillations, with maximal mixing between flavour
and mass eigenstates \cite{macro-last,sk-general}.
 This solution is strongly favoured over alternative solutions, like
 $\nmns$ oscillations \cite{macro-sterile,sk-sterile},
$\nmne$ oscillations \cite{sk-general,soudan2} and other exotic possibilities,
 such as violation of relativity
principles \cite{sk-habig,vlad-oujda,fogli,glashow9799,glashow04}, neutrino
decoherence \cite{sk-habig,sk-dip} and neutrino radiative decays 
\cite{sk-dip,eclipse}.

However, these alternative mechanisms have been considered under the hypothesis
that each one solely accounts for the observed effects. It means that
they are considered independent and mutually exclusive. Here we analyse the 
possibility of a mixed scenario: one mechanism, the mass-induced flavour 
oscillations, is considered dominant and a second mechanism is included 
in competition with the former. We studied, as sub-dominant
mechanism, neutrino flavour transitions induced by violations of relativity 
principles, i.e. violation of the Lorentz invariance (VLI) or of the 
equivalence principle (VEP).

In this mixed scenario, we assume that neutrinos can be described in 
terms of three distinct bases: flavour eigenstates, mass eigenstates and 
velocity eigenstates, the latter being characterised by different maximum 
attainable velocities (MAVs), and consider that only two families 
contribute to the atmospheric neutrino oscillations.
Thus, we may write
\be
\begin{array}{ll}
|\nu_\mu \rangle = | \nu_2^i \rangle \cos \theta_i + |\nu_3^i 
\rangle \sin \theta_i \\
|\nu_\tau \rangle = -|\nu_2^i \rangle \sin \theta_i + |\nu_3^i 
\rangle \cos \theta_i
\end{array}
\label{Eq:1}
\ee
In Eq. (\ref{Eq:1}) the index $i = m$ or $i = v$ for mass and velocity 
eigenstates, respectively.
When both mass-induced and VLI-induced neutrino oscillations are considered
simultaneously, the $\nm$ survival probability can be
expressed as~\cite{fogli,glashow9799,glashow04}
\be
P_{\nm \to \nm} = 1 - \sin^2 2 \Theta \sin^2 \Omega
\label{Eq:2}
\ee
where the global mixing angle $\Theta$ and the term $\Omega$ are given by:
\be
\begin{array}{ll}
2\Theta = \arctan (a_1/a_2) \\
\Omega = \sqrt{\left( a_1^2 + a_2^2 \right)}
\end{array}
\label{Eq:3}
\ee
The terms $a_1$ and $a_2$ in Eq. (\ref{Eq:3}) contain the relevant 
physical information
\be
\begin{array}{ll}
a_1 = 1.27 | \Dm2 \sin 2 \theta_m L/E + 2 \cdot 10^{18} \Dv \sin 2 \theta_v
LE e^{i \eta} | \\
a_2 = 1.27\left( \Dm2 \cos 2\theta_m L/E + 2 \cdot 10^{18} \Dv \cos 2
\theta_v LE \right) ~,
\end{array}
\label{Eq:4}
\ee
where the muon neutrino pathlength $L$ is expressed in km, the neutrino
energy $E$ in GeV and the oscillation parameters $\Dm2 = m^2_{\nu_3^m} -
m^2_{\nu_2^m}$ and $\Dv = v_{\nu_3^v} - v_{\nu_2^v}$ are in eV$^2$ and
$c$ units, respectively. The unconstrained phase 
$\eta$ refers to the connection between the mass and velocity eigenstates.

Note that in the limit $\Dv=0$ ($\Dm2 =0$) we recover
pure mass-induced (VLI-induced) oscillations and the functional form of the
oscillation probabilities exhibits an $L/E_\nu$ ($L \cdot E_\nu$) 
dependence on the neutrino energies and pathlengths.
Moreover, in the limiting cases of pure oscillations, the survival 
probabilities do not depend on the sign of the mixing angle and/or on 
the sign of the $\Dv$ and $\Dm2$ parameters; this is not so in the case 
of mixed oscillations, as can be seen from Eq. (\ref{Eq:4}), where the 
relative sign between the mass-induced and VLI-induced oscillation terms 
is important. The whole domain of variability of the parameters can be 
accessed with the requirements 
$\Dm2 \ge 0$, $0 \le \theta_m \le \pi/2$, $\Dv \ge 0$
and $-\pi/4 \le \theta_v \le \pi/4$.

The same formalism also applies to violation of the equivalence principle,
after substituting $\Dv/2$ with the adimensional product 
$|\phi| \Delta \gamma$; $\Delta \gamma$ is the difference of the coupling 
constants for neutrinos
of different types to the gravitational potential $\phi$ \cite{gasperini}.

\begin{figure}
\begin{center}
\mbox{\epsfig{figure=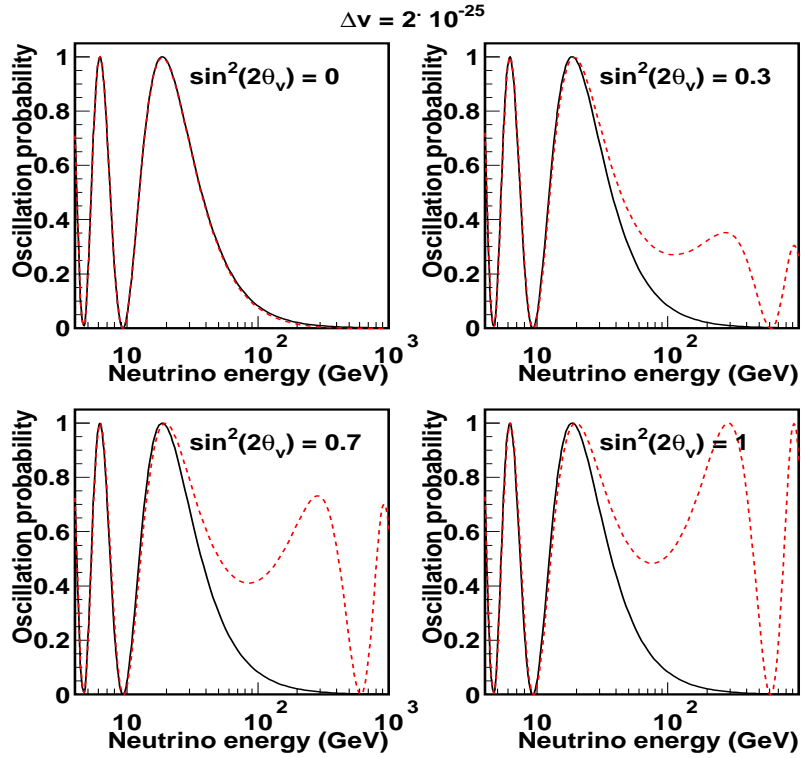,width=10.5cm,height=10cm}}
\caption {Energy dependence of the $\nmnt$ oscillation probability
for mass induced oscillations alone (continuous curves), and mass-induced 
+ VLI oscillations for $\Dv = 2 \cdot 10^{-25}$ and different values of the
$\s2t_v$ parameter (dashed curves). The
neutrino pathlength was fixed at $L = 10^4$ km.}
\label{fig:ftest}
\end{center}
\end{figure}

As shown in \cite{glashow9799,glashow04,strumia},
the most sensitive tests of VLI can be made by analysing the high energy tail
of atmospheric neutrinos at large pathlength values. As an example,
Fig. \ref{fig:ftest} shows the energy dependence of the $\nmnt$ oscillation
probability as a function of the neutrino energy, for neutrino mass-induced
oscillations alone and for both mass and VLI-induced oscillations for
$\Dv = 2 \cdot 10^{-25}$ and different values of the $\s2t_v$ 
parameter. Note the large sensitivity for large neutrino energies and 
large mixing angles.
Given the very small neutrino mass ($m_{\nu} \aprle 1$ eV), neutrinos with
energies larger than 100 GeV are extremely relativistic, with Lorentz
$\gamma$ factors larger than $10^{11}$.

In the next section we summarise the main results of the global analysis 
of MACRO neutrino data \cite{macro-last}, which used the information from 
i) the zenith angle distribution of upward-throughgoing muons, ii) from 
the energy distribution of a sub-sample of them and iii) from the low 
energy neutrino data.

In Sect. \ref{sec:VLI}, we consider the sub-sample of events in point ii) and
re-analyse their energy distribution in terms of mass-induced + VLI-induced 
mixed oscillations \cite{macro-vli}.
Since we consider mass-induced oscillations as the dominant source of 
neutrino flavour transitions, we fix the parameter $\Dm2$ and $\s2t_m$
at the values obtained from the global analysis and leave the VLI 
parameters free to vary. Finally, the constrains on $\Dm2$ and $\s2t_m$ 
parameters are partially relaxed and a multidimensional analysis is performed.

\section{MACRO results on mass-induced atmospheric neutrino oscillations}
\label{sec:MACRO}

MACRO \cite{macrodet} was a multipurpose large area detector
($\sim 10000$ m$^2$ sr acceptance for an isotropic flux)
located in the Gran Sasso underground Lab,
shielded by a minimum rock overburden of 3150 hg/cm$^2$.
The detector had global dimensions of $76.6 \times 12 \times 9.3$ m$^3$
and used limited streamer tubes and scintillation counters to detect
muons. $\nm$'s were detected via charged current interactions
$\nm + N \rightarrow \mu + X$; upward-throughgoing muons were identified 
with the streamer tube system (for tracking) and the scintillator system (for
time-of-flight measurement).
Early results on atmospheric neutrinos were published in \cite{macro-95}
and in \cite{macro-98} for the upward-throughgoing muon sample
and in \cite{macro-le} for the low energy semi-contained and
upgoing-stopping muon events. Matter effects in the $\nmns$ channel 
were presented in \cite{macro-sterile} and a global analysis of all neutrino
data in \cite{macro-last}.

In the following we shall briefly recall the analyses performed on different
samples of upgoing muon data, whose combination yielded the mass-induced
atmospheric neutrino oscillation parameters: $\Dm2 = 2.3 \cdot 10^{-3}$ 
eV$^2$ and $\s2t_m =1$.

\begin{figure}
\begin{center}
\mbox{\epsfig{figure=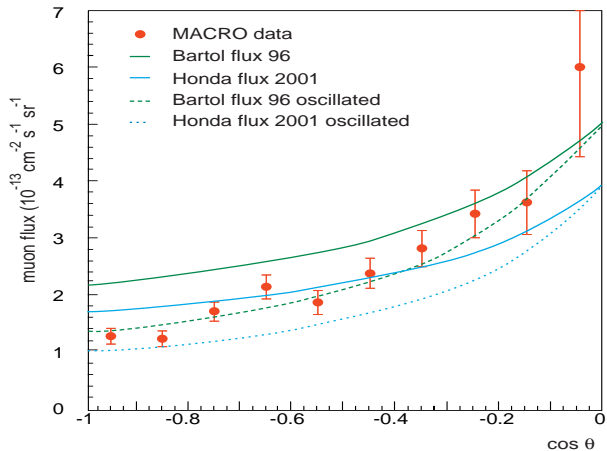,height=6cm,width=8cm}}
\caption {Comparison of the MACRO upward-throughgoing muons (black circles) 
with the predictions of the Bartol96 and of the Honda2001 MC oscillated 
and non oscillated fluxes (oscillation parameters $\Dm2 = 2.3 \cdot 10^{-3}$ 
eV$^2$ and $\s2t_m =1$).}
\label{fig:zenith}
\end{center}
\end{figure}

The first method, (already used in early analyses) is based on the study of
the zenith angle distribution of upward-throughgoing muons.
Fig. \ref{fig:zenith} shows MACRO data (black circles) compared with two 
Monte Carlo (MC) predictions: the Bartol96 \cite{bartol} flux with and 
without oscillations (the dashed and solid lines respectively) and the 
Honda2001 flux \cite{honda01}, the lower curves;
the new FLUKA MC predictions agree perfectly with the new Honda \cite{ggmg}.
The shapes of the angular distributions are very similar, but the
absolute flux values predicted by the two simulations differ by about 25\%;
more specifically, if we consider the flux extrapolation towards the
horizontal direction where no oscillations are expected, we find that 
our data are in good agreement with the Bartol96 MC.

For a subsample of 300 upward-throughgoing muons MACRO estimated the 
muon energies via multiple Coulomb scattering in the 7 horizontal 
rock absorbers in the lower apparatus \cite{mcs1,mcs2}. The energy 
estimate was obtained using the streamer tubes in drift mode, which allowed to
considerably improve the spatial resolution of the detector ($\sim 3$ mm).
The overall neutrino energy resolution was of the order of 100\%,
mainly dominated by muon energy losses in the rock below the detector
(note that $\ll E_{\mu} \rr \simeq 0.4 \ \ll E_{\nu} \rr$).
Upgoing muon neutrinos of this sample have large zenith angles
($> 120^\circ$) and the median value of neutrino pathlengths
is slightly larger than 10000 km.

\begin{figure}
\begin{center}
\mbox{\epsfig{figure=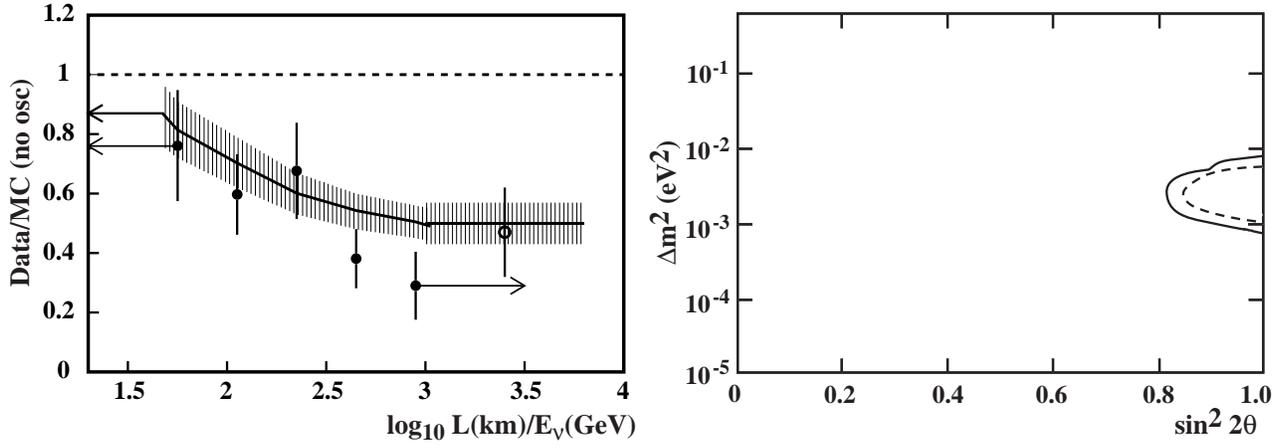,height=6cm}}
\caption {(left) Ratio Data/MC$_{\mbox{no~osc}}$ 
as a function of the estimated $L/ E_\nu$ for the upthroughgoing muon 
sample (black points). The solid line is the MC expectation assuming 
$\Dm2 = 2.3 \cdot 10^{-3}$ eV$^2$ and $\s2t = 1$. The last point 
(empty circle) is obtained from the IU sample. (right) Interpolated $90\%$ C.L.
contour plots of the allowed regions in the $\s2t - \Dm2 $ plane for the 
MACRO data using only the ratios $R_1,R_2,~R_3$ (continuous line) and adding
also the information on the absolute values $R_4,~R_5$ (dotted line).}
\label{fig:le-contour}
\end{center}
\end{figure}

Fig. \ref{fig:le-contour}a shows the ratio data/MC as a function of the 
estimated $L/E_\nu$ for the upthroughgoing muon sample. 
The black circles are data/Bartol96 MC (assuming no oscillations); the
solid line is the oscillated MC prediction for $\Dm2 = 2.3 \cdot 10^{-3}$ 
eV$^2$ and $\s2t_m =1$. The shaded region represents the simulation 
uncertainties. The last point (empty circle) is obtained
from semicontained upward going muons.

In order to reduce the effects of  systematic uncertainties in the MC 
absolute fluxes we used the following three independent ratios 
\cite{macro-last}:
\begin{enumerate}
        \item [(i)] High Energy data: vertical/horizontal ratio,
        $R_{1} = N_{vert}/N_{hor}$
        \item [(ii)] High Energy data: low energy/high energy ratio,
        $R_{2} = N_{low}/N_{high}$
        \item [(iii)] Low Energy data: 
        $R_{3} = (Data/MC)_{IU}/(Data/MC)_{ID+UGS}$
\end{enumerate}
With these ratios, the no oscillation hypothesis has a probability
$P \sim 3 \cdot 10^{-7} $ and is ruled out at the $\sim 5 \sigma$ level.
By fitting the 3 ratios  to the $\nmnt$ oscillation formulae we obtain 
$\s2t_m = 1$, $\Dm2 = 2.3 \cdot 10^{-3}$ eV$^{2}$ and the allowed 
region indicated by the solid line in Fig. \ref{fig:le-contour}b.

Using the Bartol96 flux we may add the information on the absolute 
fluxes of the
\begin{enumerate}
   \item[(iv)] High Energy data (systematic scale error of $\simeq 17 \%$): 
    $R_{4} = N_{meas}/N_{Bartol96}$.
   \item [(v)] Low Energy semicontained muons, with a systematic scale
     error of $21 \%$, $R_{5} \simeq N_{meas}/N_{Bartol96}$.
     \end{enumerate}
These informations reduce the area of the allowed region  (dashed line in 
Fig. \ref{fig:le-contour}b), do not change the best fit values and bring the
oscillation significance to $\sim 6 \sigma$.

\section{Exotic contributions to mass-induced oscillations}
\label{sec:VLI}

In order to analyse the MACRO data in terms of VLI, we used the aforementioned 
subsample of 300 upward-throughgoing muons whose energies were estimated via
Multiple Coulomb Scattering.

We used two independent and complementary analyses: one based on the $\chi^2$ 
criterion and the Feldman and Cousins prescription \cite{felcou}, and a second 
one based on the maximum likelihood technique \cite{macro-vli}.

\subsection{$\mathbf{\chi^2}$ Analysis}
\label{sub:chi} 

Following the analysis in Ref. \cite{mcs2}, we selected a low and a 
high energy sample by requiring that the reconstructed neutrino energy 
$E^{rec}_\nu$ should be
$E^{rec}_\nu < 30$ GeV and $E^{rec}_\nu > 130$ GeV.
The number of events surviving these cuts is $N_{low}$ = 49 and $N_{high}$ =
58, respectively; their median energies, estimated via Monte Carlo,
are 13 GeV and 204 GeV (assuming mass-induced oscillations).

The analysis then proceeds by fixing the neutrino mass oscillation parameters 
at the values obtained with the global analysis of all MACRO neutrino data
\cite{macro-last}:
$\overline{\Dm2} = 2.3 \cdot 10^{-3}$ eV$^2$, sin$^2$2$\overline{\theta}_m =1$.
The factor $e^{i\eta}$ for the moment is assumed to be real 
($\eta$ = 0 or $\pi$). Then, we scanned the plane of the two free parameters
($\Delta v$, $\theta_v$) using the function

\be
  \chi^2 = \sum_{i=low}^{high} \left(\frac{N_i-\alpha
    N_i^{MC}(\Dv, \theta_v; \overline{\Dm2}, \overline{\theta}_m)}{\sigma_i} 
\right)^2
\ee

\ndt where $N_i^{MC}$ is the number of events predicted by Monte Carlo,
$\alpha$ is a constant which normalises the number of Monte Carlo events
to the number of observed events and $\sigma_i$ is the overall error
comprehensive of statistical and systematic uncertainties \cite{macro-vli}.

We used the Monte Carlo simulation described in \cite{mcs2} with different
neutrino fluxes in input \cite{bartol,honda01,fluka01,newhonda}.
The largest relative percentage difference of the extreme values of 
the MC expected ratio $N_{low}/N_{high}$ is $13\%$.
However, in the evaluation of the systematic error, the main sources
of uncertainties for this ratio
(namely the primary cosmic ray spectral index and neutrino cross sections)
have been separately estimated and their effects added in quadrature
(see \cite{mcs2} for details):
in this work, we use a conservative $16\%$ theoretical systematic error
on the ratio $N_{low}/N_{high}$.
The experimental systematic error on the ratio is estimated to be $6\%$.
In the following, we show the results obtained with the
computation in \cite{newhonda}.

\begin{figure}[ht]
\begin{center}
\mbox{\epsfig{figure=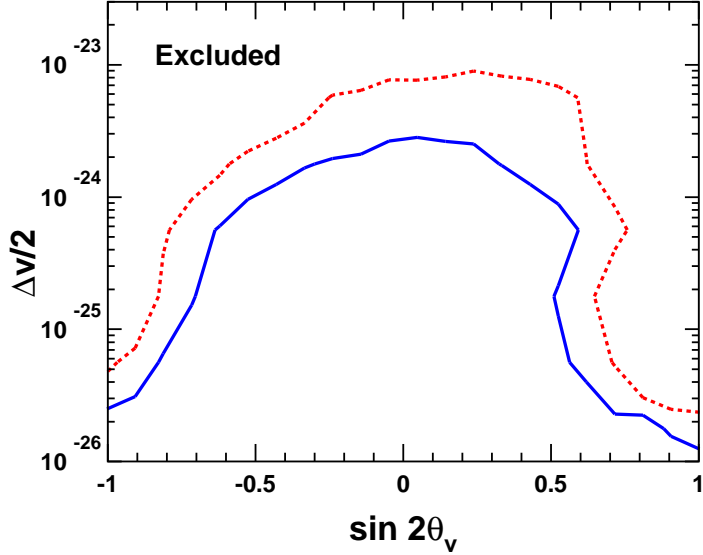,width=11cm}}
\vspace{-2cm}
\caption {$90\%$ C.L. upper limits on the Lorentz invariance violation
parameter $\Delta v$ versus sin $2\theta_v$. Standard mass-induced 
oscillations are assumed in the two-flavour $\nmnt$ approximation, with 
$\Dm2 = 2.3 \cdot 10^{-3}$ eV$^2$ and $\theta_m = \pi/4$. The dashed line 
shows the limit obtained with the same selection criteria of Ref. 
\protect\cite{mcs2} to define the low and high energy samples; the
continuous line is the final result obtained with the selection criteria
optimised for the present analysis (see text).}
\label{fig:super}
\end{center}
\end{figure}

The inclusion of the VLI effect does not improve the $\chi^2$ in any point
of the ($\Dv$, $\theta_v$) plane, compared to mass-induced oscillations
stand-alone, and proper upper limits on VLI parameters were obtained.
The $90\%$ C.L. limits on $\Dv$ and
$\theta_v$, computed with the Feldman and Cousins prescription \cite{felcou},
are shown by the dashed line in Fig. \ref{fig:super}.

The energy cuts described above (the same used in Ref. \cite{mcs2}),
were optimised for mass-induced neutrino oscillations. In order to
maximise the sensitivity of the analysis for VLI induced oscillations,
we performed a blind analysis, based only on Monte Carlo events, to
determine the energy cuts which yield the best performances. The results 
of this study
suggest the cuts $E^{rec}_\nu <$ 28 GeV and $E^{rec}_\nu >$ 142 GeV;
with these cuts the number of events in the real data are
$N^{\prime}_{low}$ = 44 events and $N^{\prime}_{high}$ = 35 events.
The limits obtained with this selection are shown in
Fig. \ref{fig:super} by the continuous line.
As expected, the limits are now more stringent than for the previous choice.

In order to understand the dependence of this result with respect to the
choice of the $\overline{\Dm2}$ parameter, we varied the
$\overline{\Dm2}$ values around the best-fit point. We found that a variation
of $\overline{\Dm2}$ of $\pm 30\%$ moves up/down the upper limit of VLI
parameters by at most a factor 2.

Finally, we computed the limit on $\Dv$ marginalized with respect to
all the other parameters left free to change inside the intervals:
$\Dm2 = \overline{\Dm2} \pm 30\%,~
\theta_m = \overline{\theta}_m \pm 20\%,~ -\pi/4 \le \theta_v \le \pi/4$
and any value of the phase $\eta$. We obtained
the $90\%$ C.L. upper limit $|\Dv| < 3 \cdot 10^{-25}$.

\subsection{Maximum Likelihood Analysis}
\label{sub:like}

A different and complementary analysis of VLI contributions to the atmospheric
neutrino oscillations was made on the MACRO muon data corresponding to
 parent neutrino energies in the range 25 GeV $\leq E_\nu \leq 75$ GeV. This 
energy region is characterised by the best energy reconstruction, and the 
number of muons satisfying this selection is 106. These events are outside 
the energy ranges used in the analysis discussed in Sect. \ref{sub:chi}, thus 
the expected sensitivity to VLI (or VEP) contributions to the atmospheric 
neutrino oscillations should be lower; on the other hand, the maximum 
likelihood technique (MLT) has the advantage
to exploit the information event-by-event (it is a bin-free approach).

\begin{figure}
\begin{center}
\mbox{\epsfig{figure=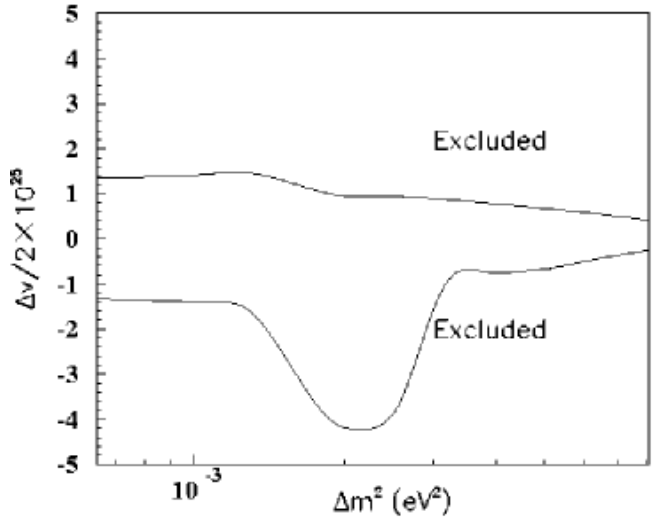,height=7cm}}
\caption {$90\%$ C.L. upper and lower limits on the $\Dv/2$  parameter, versus 
the $\Dm2$ parameter varying inside the $90\%$ C.L. MACRO global result 
\cite{macro-last}.}
\label{fig:vlad}
\end{center}
\end{figure}

Given a specific hypothesis, MLT allows to determine the set of parameters
$\mathbf{a}$ of the problem
(in our case $\mathbf{a}$ $= (\Dm2, \theta_m, \Dv, \theta_v)$)
that maximises the probability of the realization of the actual
measurements $\mathbf{x}$
(here $\mathbf{x}$ $= (E,L)$); this can be accomplished by minimising the 
negative log-likelihood function:
\be
{\mathcal{L}} = -2 \sum_{i=1}^n \ln f(\mathbf{x}_i;\mathbf{a})
\label{eq:log}
\ee
where the sum is  over the number of observed events, and
$f(\mathbf{x}_i;\mathbf{a})$ is (at least proportional to)
the probability of realization of a given event, which in our case is:
\be
f(\mathbf{x};\mathbf{a}) = \mathcal{K}(\mathbf{a}) \cdot \mathcal{P}_{MC}(\mathbf{x})
 \cdot P_{\nm \to \nm}(\mathbf{x};\mathbf{a})
\label{eq:log2}
\ee
In Eq. (\ref{eq:log2}), $\mathcal{K}(\mathbf{a})$ is a normalisation factor
meant to ensure that the integral of $f(\mathbf{x};\mathbf{a})$
over the observables $\mathbf{x}$ space is independent on
the parameters $\mathbf{a}$ (otherwise $\mathcal{L}$ would not converge),
$\mathcal{P}_{MC}(\mathbf{x})$
is the MC probability to observe the event $\mathbf{x}$ in
the no-oscillation hypothesis and $P_{\nm \nm}(\mathbf{x};\mathbf{a})$
is the $\nm$ survival probability given by Eq. (\ref{Eq:2}).

MLT has a drawback: $\mathcal{L}$ is not a true goodness of fit estimator,
as the more popular $\chi^2$ we used in the analysis in Sect. \ref{sub:chi};
at the same time, it has the advantage to be very effective when the 
experimental statistics is limited\footnote{We tested the MLT on the 
106 events sub-sample of the
data in the hypothesis of only mass-induced neutrino oscillations, obtaining
results compatible with those reported in \cite{macro-last}, based on the
full MACRO statistics.}.

We have chosen different fixed values of the
$\Dm2$ and $\s2t_m$ mass-oscillation parameters on the $90\%$ C.L.
border reported in Ref. \cite{macro-last} (the solid curve in Fig. 
\ref{fig:le-contour}b) and found the relative
$\Dv$ and $\s2t_v$ that minimise Eq. (\ref{eq:log}). Fig. \ref{fig:vlad}
shows the $90\%$ C.L. limits of the VLI parameter $\Dv/2$ versus the
assumed $\Dm2$ values.

The limit shown in Fig. \ref{fig:vlad} was obtained as a convolution of the 
$90\%$ C.L. upper limits of $|\Dv/2|$ corresponding to each chosen 
$\Dm2$ value.

\section{Conclusions}
\label{sec:conclu}

 We have searched for ``exotic'' contributions to standard mass-induced
atmospheric neutrino oscillations arising from a possible violation of Lorentz
invariance. We used a subsample of MACRO  muon events for
which an energy measurement was made via multiple Coulomb scattering.
The inclusion of VLI effects does not improve the fit to
the data, and we conclude that these effects are disfavoured even at
the sub-dominant level \cite{macro-vli}.

Two different and complementary analyses were performed on the data, both
of them yielding compatible upper limits for the VLI contribution.

The first approach uses two sub-sets of events
referred to as the low energy and the high energy samples. The mass neutrino
oscillation parameters have been fixed to the values determined in
Ref. \cite{macro-last}, 
and we mapped the evolution of the $\chi^2$ estimator
in the plane of the VLI parameters, $\Dv$ and $\s2t_v$. No $\chi^2$ 
significant improvement was found, so we applied the Feldman Cousins method
to determine $90\%$ C.L. upper limits on the VLI parameters:
$|\Dv| < 6 \cdot 10^{-24}$ at $\sin 2{\theta}_v = 0$ and
$|\Dv| < 2.5 \div 5 \cdot 10^{-26}$ at $\sin 2{\theta}_v = \pm 1$,
see Fig. \ref{fig:super}.
In terms of the parameter $\Dv$ alone (marginalization with respect
to all the other parameters), the VLI parameter bound is (at $90\%$ C.L.)
$|\Dv| < 3 \cdot 10^{-25}$.

These results may be reinterpreted in terms of $90\%$ C.L. limits of
parameters connected with violation of the equivalence principle,
giving the limit $|\phi \Delta \gamma| < 1.5 \cdot 10^{-25}$.

The second approach exploits the information contained in a data sub-set
characterised by intermediate muon energies. It is based on the maximum 
likelihood technique, and considers the mass neutrino oscillation 
parameters inside the $90\%$ border of the global result \cite{macro-last}.
The obtained $90\%$ C.L. upper and lower limits on the $\Dv$ VLI parameter
(shown in Fig. \ref{fig:vlad} versus the assumed $\Dm2$ values)
is also around $10^{-25}$.

The two analyses yielded compatible results.

The limits reported in this paper are comparable to those estimated using
SuperKamiokande and K2K data \cite{fogli}.

\newpage
\ndt {\normalsize {\bf Acknowledgements.} I would like to acknowledge the 
cooperation of the members of the MACRO 
Collaboration and of many colleagues for 
discussions and advise, in particular G. Battistoni, P. Bernardini, S. 
Cecchini, H. Dekhissi, G.L. Fogli, G. Giacomelli, S.L. Glashow, P. Lipari, 
G. Mandrioli, L. Patrizii, V. Popa, F. Ronga, M. Sioli and M. Spurio.}


\begin{thebibliography}{15}

\bibitem{solar} J.N. Bahcall et al., JHEP {\bf 0408} (2004) 016.

\bibitem{macro-98} M. Ambrosio et al., Phys. Lett. B {\bf 434} (1998) 451.

\bibitem{macro-last} M. Ambrosio et al., Eur. Phys. J. C {\bf 36} (2004) 323.

\bibitem{sk-general} Y. Fukuda et al., Phys. Rev. Lett. {\bf 81} (1998) 1562; 
Y. Ashie et al., hep-ex/0501064.

\bibitem{soudan2} M. Sanchez et al., Phys. Rev. D {\bf 68} (2003) 113004.

\bibitem{macro-sterile} M. Ambrosio et al., Phys. Lett. B {\bf 517} (2001) 59.

\bibitem{sk-sterile} S. Fukuda et al., Phys. Rev. Lett. {\bf 85} (2000) 3999.

\bibitem{sk-habig} A. Habig: in {\it Proc. 28th International Cosmic Ray 
Conferences}, Tsukuba, Japan, 2003 (Academic Press) p. 1255.

\bibitem{vlad-oujda} V. Popa and M. Rujoiu: in
{\it Proc. Cosmic Radiations: from Astronomy to Particle Physics},
Oujda, Morocco, 2001 (Eds. G. Giacomelli et al.), Kluwer Acc. Publ., NATO 
Science Series, Vol. 42, p. 181.

\bibitem{fogli} G.L. Fogli et al., Phys. Rev. D {\bf 60} (1999) 053006.
M.C. Gonzalez-Garcia and M. Maltoni, Phys. Rev. D {\bf 70} (2004) 033010.

\bibitem{glashow9799} S.R. Coleman and S.L. Glashow, Phys. Lett. B {\bf 405} 
(1997) 249; Phys. Rev. D {\bf 59} (1999) 116008.

\bibitem{glashow04} S.L. Glashow, hep-ph/0407087.

\bibitem{sk-dip} Y. Ashie et al., Phys. Rev. Lett. {\bf 93} (2004) 101801.

\bibitem{eclipse} S. Cecchini et al., Astropart. Phys. {\bf 21} (2004) 183;
Astropart. Phys. {\bf 21} (2004) 35.

\bibitem{gasperini} M. Gasperini et al., Phys. Rev. D {\bf 38} (1988) 2635.

\bibitem{strumia} A. De Gouvea et al., Nucl. Phys. B {\bf 623} (2002) 395.

\bibitem{macro-vli} G. Battistoni et al., Phys. Lett. B. {\bf 615} (2005) 14.

\bibitem{macrodet} S. Ahlen, et al., Nucl. Instr. Meth. A {\bf 324} (1993) 337.
M. Ambrosio, et al., Nucl. Instr. Meth. A {\bf 486} (2002) 663.

\bibitem{macro-95} S. Ahlen et al., Phys. Lett. B {\bf 357} (1995) 481.

\bibitem{macro-le} M. Ambrosio et al., Phys. Lett. B {\bf 478} (2000) 5.

\bibitem{bartol} V. Agrawal et al., Phys. Rev. D {\bf 53} (1996) 1314.

\bibitem{honda01} M. Honda et al., Phys. Rev. D {\bf 64} (2001) 053011.

\bibitem{mcs1} M. Ambrosio et al., Nucl. Instr. Meth. A {\bf 492} (2002) 376.

\bibitem{mcs2} M. Ambrosio et al., Phys. Lett. B {\bf 566} (2003) 35.

\bibitem{ggmg} G. Giacomelli and M. Giorgini: in {\it Proc. 7th School on 
Non-Accelerator Astroparticle Physics}, Trieste, Italy, 2004 (Eds. R.A. 
Carrigan et al.), Singapore, 2005, p. 54; hep-ex/0504002.

\bibitem{felcou} G.J. Feldman and R.D. Cousins, Phys. Rev. D {\bf 57} 
(1998) 3873.

\bibitem{fluka01} G. Battistoni et al.: in {\it Proc. 28th International 
Cosmic Ray Conferences}, Tsukuba, Japan, 2003 (Academic Press) p. 1399.

\bibitem{newhonda} M. Honda et al., Phys. Rev. D {\bf 70} (2004) 043008.

\end{thebibliography}
\end{document}